\documentclass[twocolumn]{aastex631}
\usepackage{graphicx}	
\usepackage{amsmath}	
\usepackage{booktabs}
\usepackage{xtab}
\usepackage{multirow}
\usepackage{makecell}

\shorttitle{X-ray Polarization of the SS~433 Eastern Jet}
\shortauthors{Fan et al.}

\begin{document}

\title{SS~433 Eastern Jet Turbulence Revealed by Spatially-Resolved \textit{IXPE} Data Analysis}

\author[0009-0009-2549-1161]{Ningyue Fan}
\affiliation{Department of Physics, Stanford University, Stanford, CA 94305, USA}
\affiliation{Kavli Institute for Particle Astrophysics and Cosmology, Stanford University, Stanford, CA 94305, USA}

\author[0000-0002-6401-778X]{Jack T. Dinsmore}
\affiliation{Department of Physics, Stanford University, Stanford, CA 94305, USA}
\affiliation{Kavli Institute for Particle Astrophysics and Cosmology, Stanford University, Stanford, CA 94305, USA}

\author[0000-0001-6711-3286]{Roger W. Romani}
\affiliation{Department of Physics, Stanford University, Stanford, CA 94305, USA}
\affiliation{Kavli Institute for Particle Astrophysics and Cosmology, Stanford University, Stanford, CA 94305, USA}

\begin{abstract}
The  extended X-ray jet (EXJ) of SS~433/W50 shows hard X-ray and very-high-energy $\gamma$-ray emission, indicating strong particle acceleration, likely sensitive to the EXJ magnetic field structure and turbulence. To study this field, we re-analyze the 2023 \textit{Imaging X-ray Polarimetry Explorer} (\textit{IXPE}) observation of the eastern EXJ using the publicly available polarization fitting package \texttt{LeakageLib}. Compared with standard \texttt{PCUBE} and spectropolarimetric analyses, this approach improves the polarization constraints and enables spatially resolved polarization measurements. Dividing the EXJ into head, spine, and sheath regions with independent polarization properties, we find significant ($>3\sigma$) improvement over a single-zone model. The sheath exhibits lower polarization than the head and spine, while the electric vector position angles are consistent across all regions, indicating magnetic fields aligned with the jet axis. The measurements imply enhanced magnetic turbulence at the jet boundary, potentially driven by Kelvin–Helmholtz instabilities.

\end{abstract}




\section{Introduction}
W50, also known as the ``Manatee nebula,'' is one of the largest known Galactic supernova remnants with a linear extent of $\sim 2^{\circ} \times 1^{\circ}$ (East-West $\times$ North-South), corresponding to $\sim 200 \rm pc \times 100\rm pc$ \citep{Sunyaev2025_erosita} at the radio imaging distance of $d = 5.5 \rm \pm 0.2kpc$ \citep{blundell2004}. The nebula houses a compact object \citep[likely a black hole][]{Blundell2008_circmbinary_velocity,cherepashchuk2025} binary, SS~433, with a supercritical accretion disk \citep{fabrika2004}, and X-ray/optical/radio jets launched along a precessing axis at $\sim0.26~c$ \citep{Migliari2005_extended_iron,Sakai2025_Chandra_arcsec,Waisberg2019_GRAVITY_Xshooter,Jeffrey2016_bolide_motion}. The precession cone has a half-opening angle of $\sim20^{\circ}$ \citep{Gies2002_spectral_components}, a period of $\sim162~$days, and axis inclined by $78.9^{\circ}$ to the line of sight \citep{Margon1984}.

X-ray images show that W50 is filled with diffuse thermal emission as well as hard spectrum non-thermal emission from both the eastern and western jets (EXJs) \citep{Sunyaev2025_erosita}. The EXJs start $\sim25$\,pc from the central compact object and extend towards the outer regions of W50. Detailed imaging and spectral analyses with \textit{XMM-Newton} and \textit{NuSTAR} show that the EXJ can be separated into head, cone and surrounding diffuse regions \citep{Safi-Harb_2022}. The head features a power-law spectrum with a photon index $\Gamma\sim1.6$, while the spectrum softens in the outer portions of the EXJ.

Assuming that the EXJ head emission is synchrotron radiation with an equipartition magnetic field strength of $\sim12~\mathrm{\mu G}$, the electron energy must reach 100s of TeV \citep{Safi-Harb_2022}. This is also indicated by the GeV$-$several hundred TeV (likely Compton-produced) $\gamma$-rays detected from the W50 lobes by \textit{Fermi}-LAT, \textit{VERITAS}, \textit{HAWC}, \textit{H.E.S.S.}, and \textit{LHAASO} \citep{Bordas2015_fermi,VERITAS_2026,HAWC2024,HESS2024,LHAASO2024}. To produce such high electron energies, a recollimation shock at the EXJ head is often invoked as the site of particle acceleration. The magnetic field and field turbulence within the EXJ are crucial ingredients for understanding the particle acceleration at the recollimation shock and the downstream transport of accelerated particles where they interact with the surrounding supernova remnant. The \textit{Imaging X-ray Polarimetry Explorer} \citep[\textit{IXPE},][]{2022JATIS...8b6002W} can constrain these fields, with the mean magnetic field orientation inferred from the measured electric vector polarization angle (EVPA), and the polarization degree (PD) probing the magnetic field ordering and turbulence level. 

\textit{IXPE} observed the eastern EXJ in 2023 and the western EXJ in April and October-November of 2025. It is also scheduled to observe a downstream region of the eastern jet in late 2026. Analysis of the eastern EXJ data \citep{Kaaret2024_ixpe} has revealed an EVPA consistent with $0^{\circ}$, indicating a mean magnetic field parallel with the jet axis, with an estimated PD of 39\%$-$77\%. These measurements, made with standard \texttt{PCUBE} \citep{ixpeobssim} and spectropolarimetric fitting, are however limited by the faintness of the EXJ emission and by the substantial background. Here we apply the advanced \textit{IXPE} analysis techniques developed in \cite{Dinsmore2025_advanced_weights} to obtain improved polarization constraints on the eastern EXJ. These techniques are particularly helpful for low surface brightness sources. The paper is organized as follows: Section~\ref{section2} describes the data analysis pipeline, including how the data is processed and how the spatial and spectral weights are obtained. Section~\ref{result} presents the spatially resolved polarization measurements of the EXJ head, spine and sheath, and Section~\ref{discussion} discusses physical interpretations of our results.

\section{Observation and Data Analysis}\label{section2}

\subsection{\textit{IXPE} Observation and Data Processing}\label{data analysis}

We treat the 2023 \textit{IXPE} data for the eastern jet of SS~433 (ObsID: 02006799), observed from April 19 to May 9 with a gap on May 2, for a total exposure $\sim890$~ks. The \textit{IXPE} field-of-view (FoV) was centered on the onset of the Eastern EXJ at (R.A., Decl.) = $(19^{\rm h}13^{\rm m}06.5^{\rm s}, +04^\circ56'11'')$, around 25~pc away from SS~433 itself \citep{Sunyaev2025_erosita}.

As shown in \cite{Kaaret2024_ixpe}, some data pruning is required before polarization fitting. To reduce bright solar flares, we excise periods when the count rate exceeds the mean rate by more than 3 standard deviations \citep{Kaaret2024_ixpe,Forsblom2023}. In practice, solar-associated photon and particle events are present even outside these strong flares; this residual background will be modeled later (see subsection~\ref{specral weight subsection}). We identify background cosmic ray (CR) particles with the particle-weighting scheme \texttt{leakagelib\_bkg} described in \cite{Dinsmore2025_advanced_weights}. Because the source is extended, very faint and background-dominated, these steps are essential for reliable polarization measurements. For the remaining (largely) photon events we use event parameters reconstructed from both the standard ``Moments'' (Mom) pipeline \citep{Bellazzini2003_moments} and the neural network (NN) pipeline developed by \cite{Peirson2021}.

To improve on \texttt{PCUBE} or spectropolarimetric fitting, we use the maximum likelihood method polarization fitting of the \texttt{LeakageLib} package\footnote{\url{https://jack-dinsmore.github.io/leakagelib/index.html}} \citep{Dinsmore2025_advanced_weights,Dinsmore2025_leakagelib}. Since {\it Chandra} and {\it XMM} measurements provide the spatial and spectral distributions of the source and background, \texttt{LeakageLib} can perform an event-by-event source-weighted maximum-likelihood fit to the polarization parameters. The spatial and spectral weights used are described in detail in the following subsections. Importantly, the fit likelihood allows us to test the significance of our results. If the difference between the log-likelihood of two models is $\Delta\ln L$, in certain limits $-2\Delta\ln L$ is $\chi^2$ distributed with $k$ degrees of freedom \citep[Wilks' theorem, see e.g.,][]{wilks2015elementary}. Here $k$ is the difference in the number of free parameters between the two nested models.
The $p$-value quantifying the probability that the improved likelihood purely arises from statistical fluctuations is $p=\hat\Gamma(k/2,-\Delta\ln L)$, where $\hat\Gamma$ is the normalized upper incomplete $\Gamma$-function. The $\sigma$ significance can then be calculated as $\sigma = \sqrt{2}\times {\rm erfc}^{-1}(p)$, where ${\rm erfc}^{-1}$ is the inverse complementary error function. Since the maximum-likelihood fitting is performed directly in the Stokes \(q\)-\(u\) space instead of the PD-EVPA space, the best-fit \(q\)-\(u\) values and the likelihood used to evaluate the fit difference significance are not subject to the Ricean bias associated with the positive-definite polarization degree.

\subsection{Spatial Weights}\label{spatial weight subsection}
\begin{figure*}
    \centering
    \includegraphics[width=0.8\linewidth]{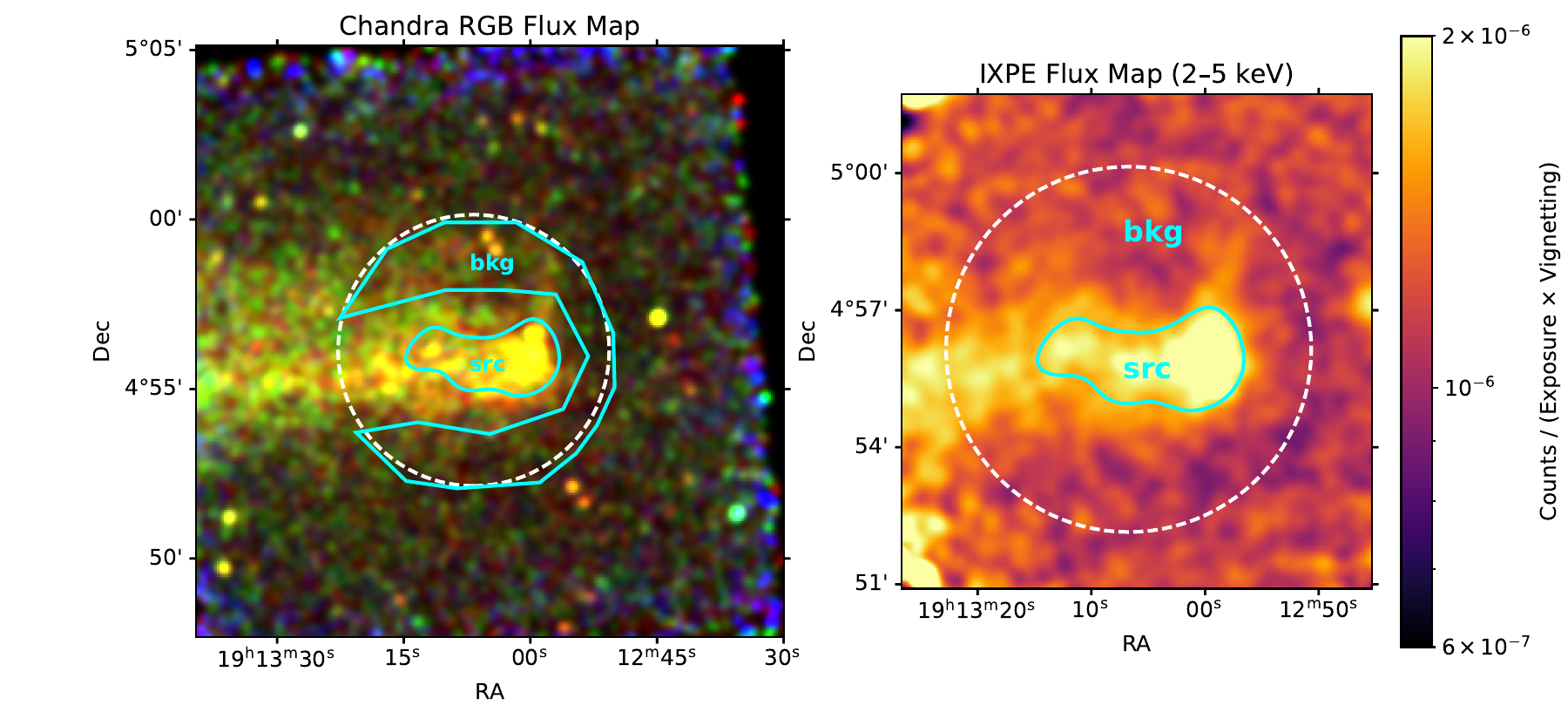}
    \includegraphics[width=0.8\linewidth]{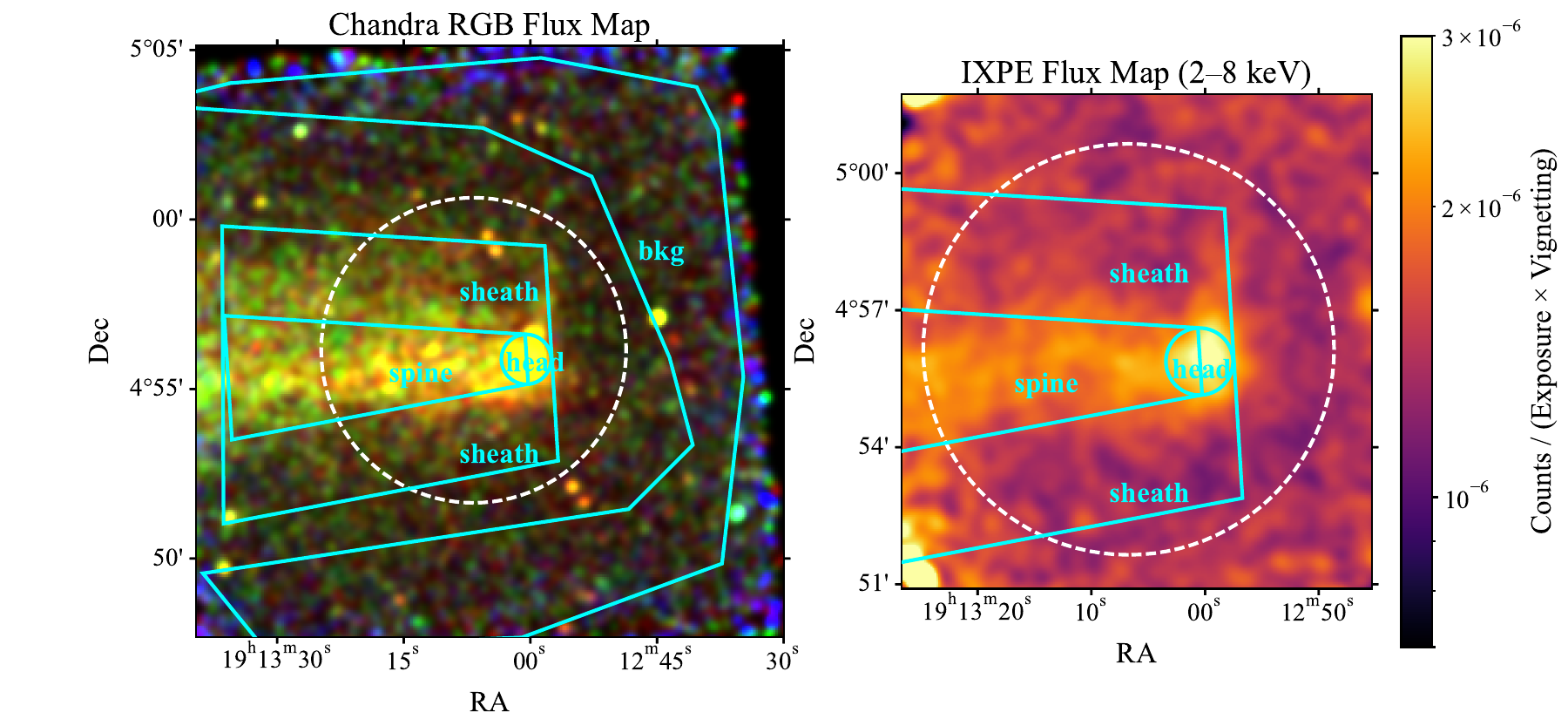}
    \caption{Analysis regions shown on the \textit{Chandra} and \textit{IXPE} flux maps. Left panels: \textit{Chandra} exposure-corrected flux maps. Red, green and blue  correspond to the soft (0.5-1.2~keV), medium (1.2-2.0~keV) and hard (2.0-7.0~keV) bands. Right panels: the \textit{IXPE} exposure-corrected flux maps. The upper right panel is the 2-5~keV flux map while the lower right panel is the 2-8~keV flux map. The cyan solid lines indicate different source regions and a \textit{Chandra} background region, while the white dashed lines indicate our IXPE FoV. The upper panels show the one-region fit and FoV choice \cite[$240''$ in radius, similar to][]{Kaaret2024_ixpe}, while the lower panels show the 3-region fit and FoV choice ($270''$ in radius).}
    \label{chandra_ixpe}
\end{figure*}

\texttt{LeakageLib} requires a model for the true flux distribution of the source in order to correctly apply spatial weights without incurring bias from polarization leakage. Due to \textit{Chandra}'s exceptionally tight $\sim0.5''$ on-axis resolution, and because the EXJ motions are small compared to the \textit{IXPE} PSF as revealed by \textit{Chandra} observations spanning 20 years \citep{Tsuji2025_proper_motion_head}, we use archival \textit{Chandra} observations\footnote{These observations are available through this Chandra Data Collection DOI: \url{https://doi.org/10.25574/cdc.692}} (ObsIDs 26533, 27064 and 29477, which overlap with the  \textit{IXPE} FoV) to model the source's true flux distribution. These observations were reprocessed with the \texttt{chandra\_repro} script\footnote{\url{https://cxc.cfa.harvard.edu/ciao/ahelp/chandra\_repro.html}} in \texttt{CIAO}, using calibration database v4.12.3 and merged with \texttt{merge\_obs}\footnote{\url{https://cxc.cfa.harvard.edu/ciao/ahelp/merge\_obs.html}}. The resulting exposure-corrected flux maps are shown in the left panels of Fig.~\ref{chandra_ixpe}.

Finally, we translate the maps to IXPE's band by subtracting the \textit{Chandra} flux inside the background region identified in the upper left panel of Fig.~\ref{chandra_ixpe}. After dividing the result by the \textit{Chandra} auxiliary response file (ARF) as output by \texttt{mkwarf} and multiplying by the IXPE ARF as output by \texttt{ixpecalcarf}, the result is an IXPE-band prediction for the distribution of source photons. We created a publicly accessible tool \texttt{leakagelib\_cxo}\footnote{\url{https://jack-dinsmore.github.io/leakagelib/examples/fit-prepare-extended.html}} to perform these steps given the output of \texttt{merge\_obs}. This tool is documented and included in the new release of \texttt{LeakageLib}.

We define three different choices for source regions, deriving associated spatial weight maps. These are shown as cyan lines in Fig.~\ref{chandra_ixpe}. The first, (C11 hereafter) is similar to the ``C11'' flux contour fitted in \cite{Kaaret2024_ixpe} and defines the brightest portion of the eastern EXJ.
The second (hereafter `separate' model) uses the \textit{Chandra} image morphology and spectral variations to divide the eastern EXJ into three regions which could have different physical nature and hence varying polarization: the `head,' a small, bright region at the jet onset, plausibly representing the SS~433 jet recollimation shock, the `spine,' the bright ridge or core of the EXJ extending from the head and the `sheath,' a lower surface-brightness region bracketing the spine.
The resulting flux ratios of the 3 regions are $F_\mathrm{head}:F_\mathrm{spine}:F_\mathrm{sheath}=1:2.53:2.66$.

For our third region choice (hereafter `overlapping' model), we acknowledge that these are 3D structures and that the spine may be embedded in a surrounding sheath. We keep the head region as a spatial aperture, but adopt a simple geometrical model to define a relative probability of spine/sheath/background origin. To formulate this, we first note that in the region covered by \textit{IXPE} the combined spine/sheath flux does not decrease significantly down-stream, nor is there evidence for spectral softening (Fig.~5 of \cite{Safi-Harb_2022} shows a constant hardness ratio in this zone, although the structure does appear softer further down-stream). Thus we infer that the spine and sheath do not experience strong cooling in the \textit{IXPE}-studied zone. The downstream broadening of the EXJ seen in the \textit{XMM-Newton} and \textit{Chandra} images, together with the decrease in surface brightness, is therefore consistent with the same X-ray-emitting flow broadening downstream, rather than the addition of a distinct emission component. Assuming Gaussian cross sections for the two components and linear expansion along the jet length, we then model the combined spine/sheath as the superposition of two Gaussians, where the two amplitudes and the (linearly increasing) half-widths are fit to the \textit{Chandra} image. Specifically, the spine and sheath Gaussian full width at half maxima (FWHM) are $(90+0.24d)''$, and $(400+0.24d)''$, where $d$ is the projected distance from the center of the head in arcsec, and the coefficient $0.24=2\times\tan7^{\circ}$ corresponds to a half opening angle of $7^{\circ}$. Thus at the sky position of each \textit{IXPE} event we can define the local amplitudes of each Gaussian component $A_{\rm sp}$ and $A_{\rm sh}$ and the spatial weights for the components from $I_{\rm{sp}}=I_{\rm{tot}}\frac{A_{\rm{sp}}}{A_{\rm{sp}}+A_{\rm{sh}}}$, and $I_{\rm{sh}}=I_{\rm{tot}}\frac{A_{\rm{sh}}}{A_{\rm{sp}}+A_{\rm{sh}}}$. This results in flux ratios $F_\mathrm{head}:F_\mathrm{spine}:F_\mathrm{sheath}=1:2.17:3.02$.

Note that \texttt{Leakagelib} identifies the background flux and polarization using weights alone, so a background aperture is not necessary for the polarization fit. The background flux and polarization are appropriately removed when reporting the source polarization, and uncertainties are propagated. \texttt{LeakageLib} does assume that the background flux and polarization are uniform across each detector's image. We checked this assumption by confirming that the fit results do not significantly change if we enlarge the FoV and include more background photons.


\subsection{Spectral Weights}\label{specral weight subsection}

\begin{figure*}
    \centering
    \includegraphics[width=0.48\linewidth]{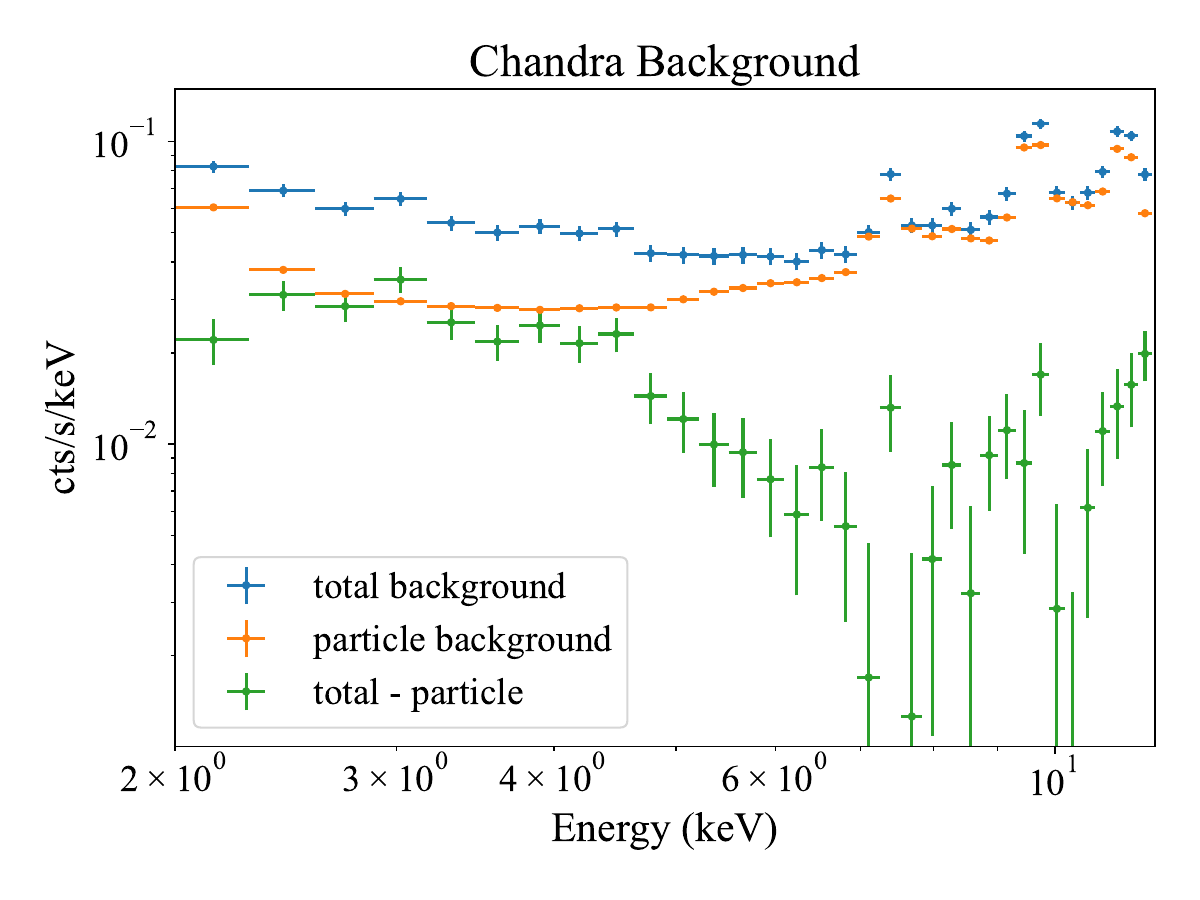}
    \includegraphics[width=0.45\linewidth]{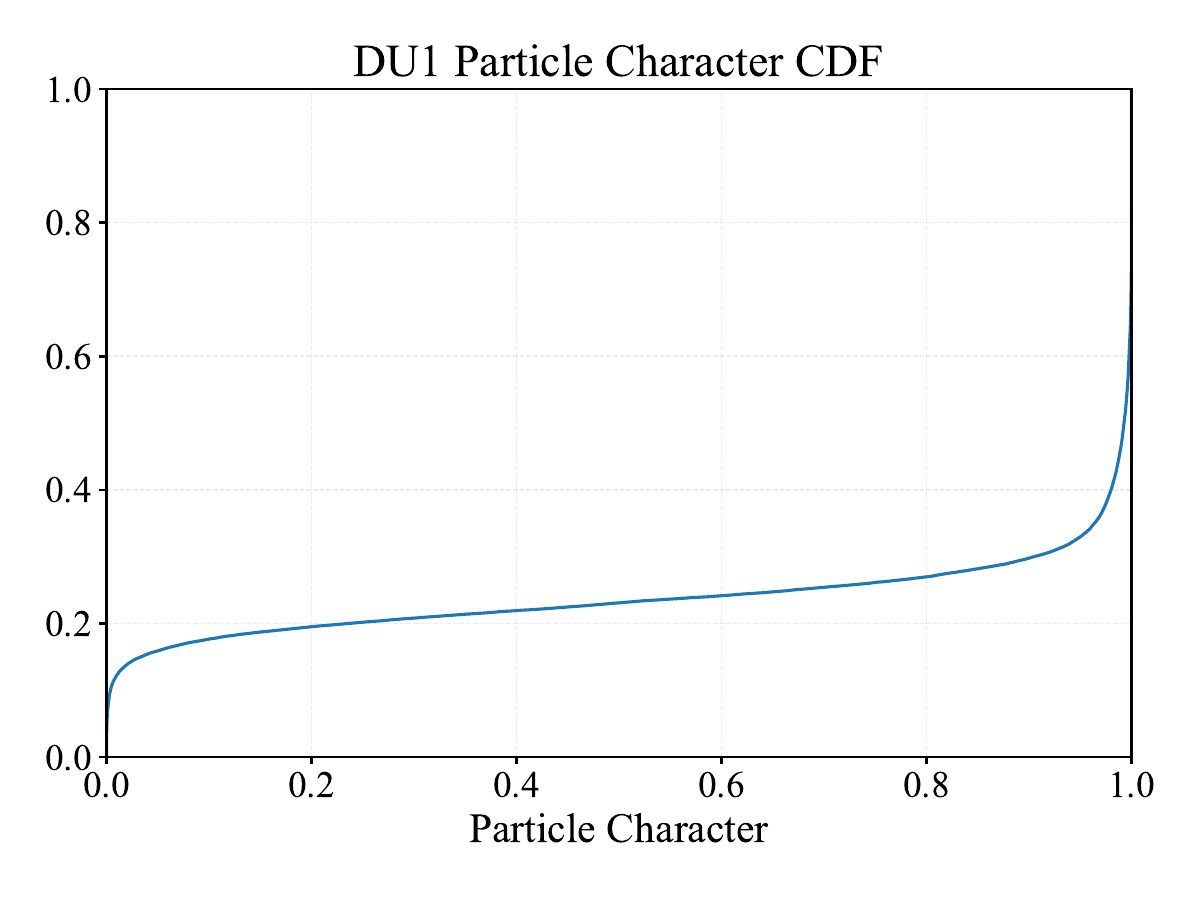}
    \includegraphics[width=0.5\linewidth]{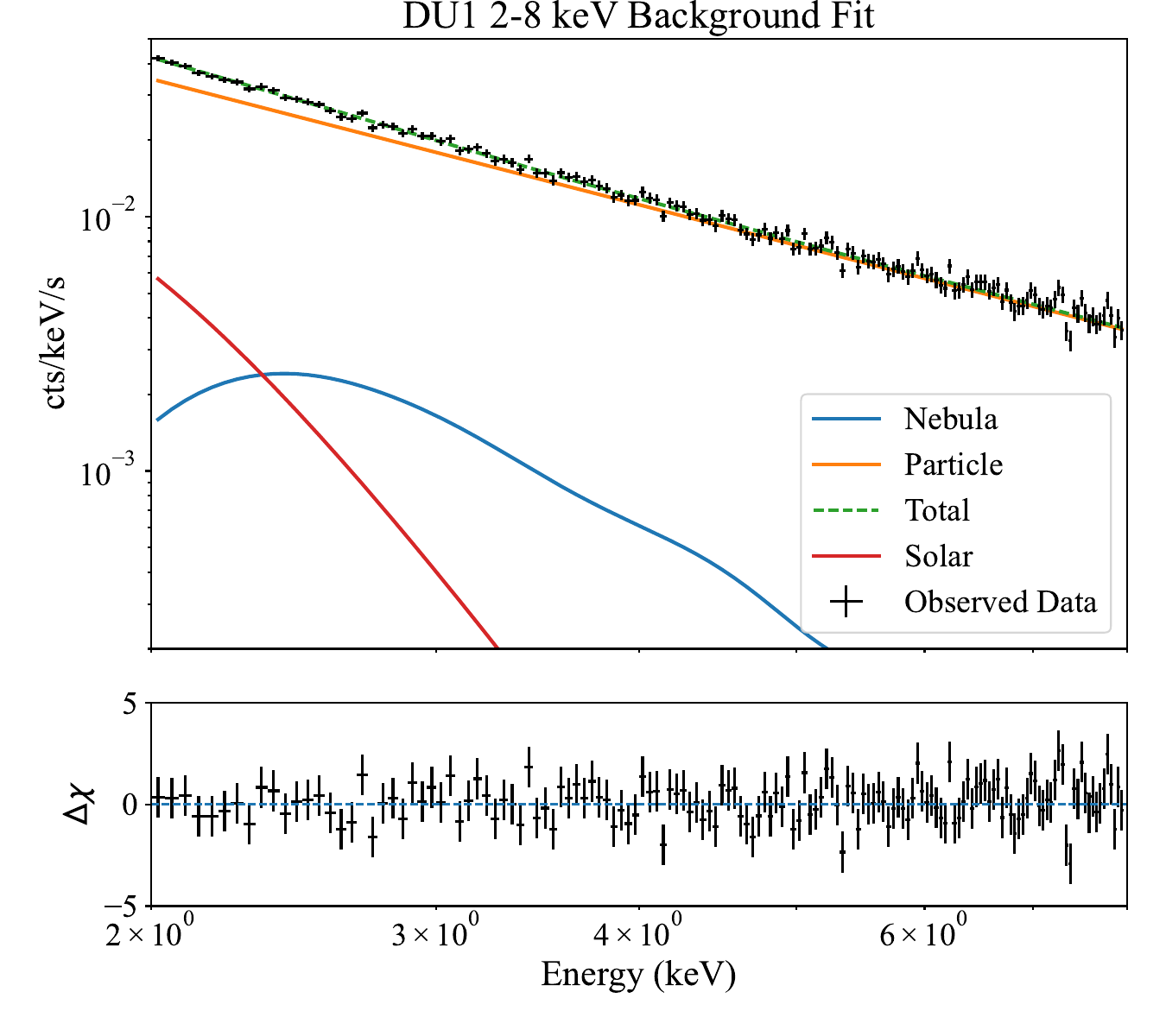}
\caption{Separation of background spectral components, using DU1 as an example. The upper-left panel shows the \textit{Chandra} total, particle and particle-subtracted background spectra. The upper-right panel shows the cumulative distribution function (CDF) of the  3-8~keV particle character predicted by \texttt{leakagelib\_bkg}, from which we derive the fraction of particles among all the events. Lower panel: The data, model and residuals obtained from fitting 2-8~keV data with particle background, astrophysical (nebula) background and a soft power-law component likely related to the Sun. The 2-8~keV background data can be well fit with those three components.}
    \label{spectral_weights}
\end{figure*}

The spatial weight maps give the probability density functions (PDFs) at event positions; corresponding spectral weights give the PDF of the event energies. Again the background-subtracted \textit{Chandra} data provide the best measurements of the source. We use \texttt{CIAO} tool \texttt{specextract}\footnote{\url{https://cxc.cfa.harvard.edu/ciao/ahelp/specextract.html}} to extract the source spectra and fit these with simple absorbed power-law models with \texttt{xspec} \citep{Xspec_Arnaud}. We find that the power-law indices for the C11 region, the head, the spine and the sheath are all consistent within $1~\sigma$. Therefore, we fit all the source regions jointly and use the resulting power-index $\Gamma=1.52\pm0.12$ in the \texttt{LeakageLib} polarization fitting. We use this spectrum to help distinguish source photons from background.

The \textit{IXPE} background contains three components: the astrophysical photon background (largely from the supernova remnant), the particle background, and the soft photon background. The particle background comes from primary and secondary cosmic rays and can produce ionization tracks resembling those generated by X-ray photoelectrons \citep{xie2021_particle_background,DiMarco2023}. The soft photon background varies among detector units (DUs) and seems correlated with solar illumination. We extract spectra for all three backgrounds. These weights provide improved source-background discrimination and allow us to fit the full 2-8~keV energy range and full $4.5'$ radius FoV, to maximize use of the available data. While we can measure the astrophysical background spectrum with \textit{Chandra}, the particle and soft photon backgrounds must be extracted from the \textit{IXPE} data.

To extract the astrophysical background photon spectrum, we use the large background region shown in the lower left panel of Fig.~\ref{chandra_ixpe}. We must, of course, first exclude the particle component in the \textit{Chandra} spectra. We use the \texttt{blanksky}\footnote{\url{https://cxc.harvard.edu/ciao/threads/acisbackground/\#acisbg}} template in \texttt{CIAO} to model the particle spectrum and flux. The total, particle and particle-subtracted backgrounds are shown in the upper-left panel of Fig.~\ref{spectral_weights}. We confirmed that this background spectrum is consistent with the spectrum in the smaller \textit{IXPE} FoV. We next map the astrophysical background to \textit{IXPE} using the \texttt{xspec} tool \texttt{fakeit}\footnote{\url{https://heasarc.gsfc.nasa.gov/docs/software/xspec/manual/node98.html\#fakeit}} and the \textit{IXPE} ARF and redistribution matrix file (RMF)\footnote{\url{https://heasarc.gsfc.nasa.gov/FTP/caldb/data/ixpe/gpd/cpf/rmf/}}. The simulated \textit{IXPE} astrophysical background is shown as the blue solid lines in the lower panel of Fig.~\ref{spectral_weights}.

To measure the IXPE particle spectrum, we extract the 3-8~keV \textit{IXPE} events and fit a power-law particle component plus the predicted astrophysical photon background. The distribution of particle characters\footnote{Particle characters are the \texttt{leakagelib\_bkg}-assigned probabilities for each track to be a particle assuming the observation contains equal parts particles and photons. This assumption can be removed, converting a particle character into a true particle probability, as discussed by \cite{Dinsmore2025_advanced_weights}.} extracted by \texttt{leakagelib\_bkg} (see the upper-right panel of Fig.~\ref{spectral_weights}) allows us to predict the amplitude of the particle component by maximizing the \texttt{LeakageLib} likelihood as a function of the flux of the particle component. The index of the particle spectrum is now the only free parameter. We find $\Gamma_{\mathrm{pr}}=1.64\pm0.02$, consistent in all three DUs.

Finally, the soft \textit{IXPE} photon excess can be fit to the lower energy events. A very soft power-law component ($\Gamma_\mathrm{soft}>6$, varying slightly between DUs) provides a good description of this excess (red solid line in Fig.~\ref{spectral_weights}).

\section{Results}\label{result}
First we check whether the \texttt{LeakageLib} fitting results are consistent with those in \cite{Kaaret2024_ixpe}. Note that their \texttt{PCUBE} fitting assumes an unpolarized background and that the observed C11 PD must be corrected for a large estimated background contamination. Thus it is better to compare with their spectropolarimetric measurement. To that end, we restrict to 2-5~keV and use the Mom event parameters as did \cite{Kaaret2024_ixpe}, but we employ all the weights for the polarized C11 source and background regions. The resulting fit PD is $49\pm11\%$ with EVPA $1.2\pm6.7^\circ$, quite consistent with the \cite{Kaaret2024_ixpe} result (PD=$52\pm14\%$, EVPA=$-3.7\pm7.8^\circ$), with slightly smaller errors ($1\,\sigma$ uncertainties are quoted throughout). This is a substantial precision improvement compared to simple aperture polarimetry techniques \citep[e.g.~\texttt{PCUBE}]{kislat2015analyzing}, which, corrected for background dilution, give a PD uncertainty of 19\%. Compared to the aperture polarimetry result, the weighted analysis shrinks the area of the 1$\sigma$ $QU$ uncertainty contour by a factor of 2.7, as has been seen in other \texttt{LeakageLib} analyses \citep{Dinsmore2026_lighthouse, Dinsmore2025_advanced_weights}.

If instead we use the NN-reconstructed parameters of the 2-5~keV events, the C11 aperture fitting gives PD=$40\pm9\%$ and EVPA=$1.6\pm6.6^\circ$. As expected, the errors decrease; the central value for the PD also decreases but lies within $1~\sigma$ of the spectropolarimetric results.

\begin{figure*}
    \centering
    \includegraphics[width=0.3\linewidth]{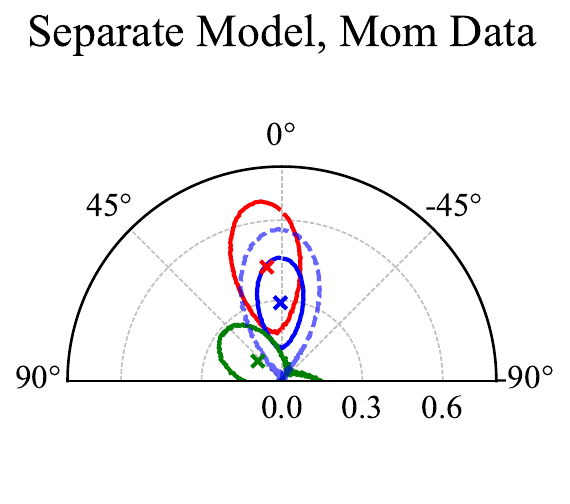}
    \includegraphics[width=0.3\linewidth]{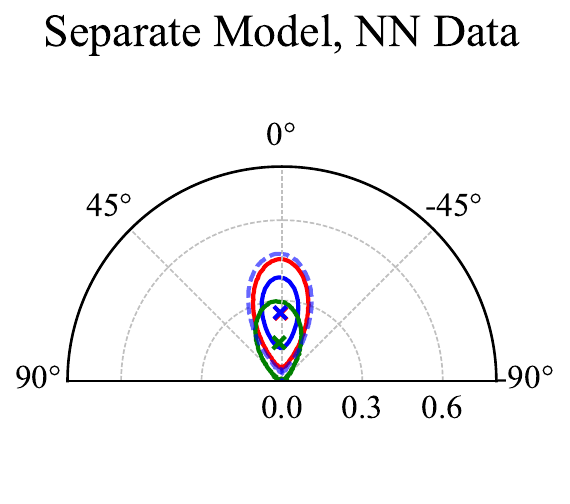}\\
    \includegraphics[width=0.3\linewidth]{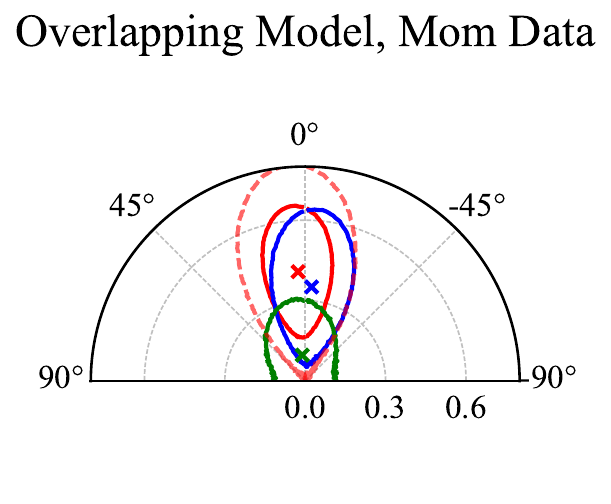}
    \includegraphics[width=0.3\linewidth]{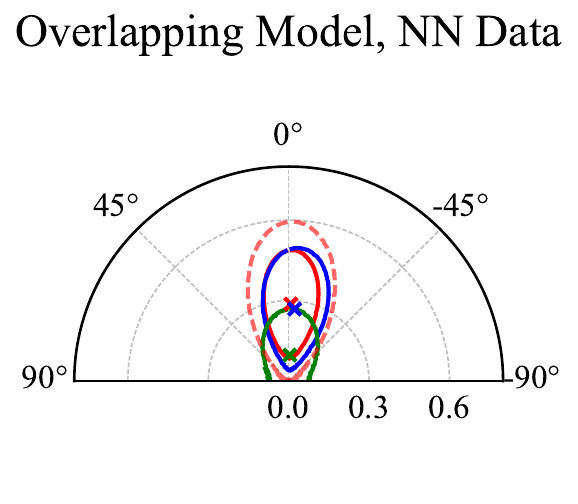}
\caption{The EVPA/PD contours of different fits. The best fit values are marked with crosses, and the uncertainties are shown at $1~\sigma$ level with solid lines. For each fit, the dashed line shows the $2~\sigma$ range of the best-constrained component. Red, blue and green lines indicate the head, spine and sheath fitting results, respectively. The EVPAs are measured positive east of north, with $0^\circ$ corresponding to north and $90^\circ$ to east. The jet axis in Fig.~\ref{chandra_ixpe} is therefore at $\sim90^\circ$. The different $\sigma$-level contours are drawn from histograms of the PD-EVPA distribution, which are weighted by 1/PD to impose a uniform-PD prior and properly represent the parameter space near PD=0.}
    \label{papad_contour}
\end{figure*}

However, our particle weights afford us improved sensitivity to photons in the higher energy 5-8 keV band, and our use of spatial weights allows us to fit several adjacent or even overlapping regions. Thus we are encouraged to attempt the multi-zone fits with the full 2-8 keV nominal IXPE band. The results are summarized in Table~\ref{result_table} and the polarization contour plots are shown in Fig.~\ref{papad_contour}. Note that the EVPAs of all components are consistent with perpendicular to the EXJ axis at $\sim94^\circ$ (i.e.~magnetic field aligned with the jet), so the single zone C11 fitting did not suffer PD dilution from zone mixing. Nevertheless, the multi-zone analysis does detect polarization differences between the regions, improving the model fit at 3.2$\sigma$ net significance. This significance is derived from the likelihood increase $\Delta {\log} L$ with respect to a single-zone model. The PD variation suggests a weakly polarized sheath zone compared to the head or spine. The lower signal-to-noise ratio in the sheath leads to a larger uncertainty in its polarization measurement, but since we properly model the background in all three regions, this does not systematically bias the measurement toward a lower PD in the sheath. Comparing the NN and Mom results, the former show significantly smaller $\sigma_{\rm PD}$ but also show smaller central values and so are, in this sense, conservative.

Note also that while the `overlapping' model with superimposed Gaussian components does a better job of matching the data (larger $\Delta {\log} L$), it does give somewhat larger errors for the spine and sheath polarization parameters; this is understandable since these regions overlap, which induces a high correlation ($\sim 55\%$ between the sheath and spine Stokes coefficient). Nevertheless, the overall PD trend with no significant EVPA shift is clear, and the \texttt{LeakageLib} three-zone analysis provides more complete exploitation of the data than a single component model.

Because the measured EVPAs are similar, we also attempted fits with all three zones fit to a common EVPA. We find that the PD fit values are in all cases consistent with the results in Table~\ref{result_table}. The spine and sheath PD errors decrease slightly, while the 
head q/u values correlate more strongly with those of the other zones and its PD error shows a slight increase. The overall multizone significances increase, to $>3.9\sigma$.

\begin{table*}
\centering
\begin{tabular}{c c c}
\hline
 &
\textbf{Moments data} &
\textbf{NN data} \\
\textbf{Total Counts} & 167499 & 160701\\
\hline

\textbf{Separate Model} & & \\

Head &
\makecell[c]{
PD=$43\pm15\%$, EVPA=$7.5\pm10.4^\circ$\\
Detection: $1.57~\sigma$\\
Counts: 706
}
&
\makecell[c]{
PD=$25\pm12\%$, EVPA=$0.9\pm14^\circ$\\
Detection: $1.09~\sigma$\\
Counts: 703
}
\\
\\
Spine &
\makecell[c]{
PD=$29\pm11\%$, EVPA=$1.1\pm10.3^\circ$\\
Detection: $1.60~\sigma$\\
Counts: 1786
}
&
\makecell[c]{
PD=$26\pm8\%$, EVPA=$1.5\pm9.4^\circ$\\
Detection: $1.76~\sigma$\\
Counts: 1780
}
\\
\\
Sheath &
\makecell[c]{
PD=$12\pm14\%$, EVPA=$50\pm35^\circ$\\
Detection: $0.59~\sigma$\\
Counts: 1878
}
&
\makecell[c]{
PD=$14\pm12\%$, EVPA=$4.2\pm23^\circ$\\
Detection: $0.89~\sigma$\\
Counts: 1871
}
\\
\\
Photon Background & 
\makecell[c]{
DU1: PD=$3\pm6\%$, EVPA=$32\pm46^\circ$\\
DU2: PD=$5\pm6\%$, EVPA=$-43\pm36^\circ$\\
DU3: PD=$10\pm6\%$, EVPA=$-37\pm8^\circ$\\
Total Counts: 32991
} & 
\makecell[c]{
DU1: PD=$4\pm4\%$, EVPA=$-89\pm25^\circ$\\
DU2: PD=$3\pm4\%$, EVPA=$-70\pm42^\circ$\\
DU3: PD=$9\pm4\%$, EVPA=$-5\pm12^\circ$\\
Total Counts: 32092
}\\
Particle Background &
\makecell[c]{
DU1: PD=$1.6\pm0.7\%$, EVPA=$-80\pm12^\circ$\\
DU2: PD=$0.4\pm0.7\%$, EVPA=$28\pm45^\circ$\\
DU3: PD=$1.5\pm0.7\%$, EVPA=$68\pm13^\circ$\\
Total Counts: 130138
}
& 
\makecell[c]{
DU1: PD=$2.7\pm0.8\%$, EVPA=$-76\pm8^\circ$\\
DU2: PD=$1.5\pm0.8\%$, EVPA=$24\pm15^\circ$\\
DU3: PD=$4.4\pm0.8\%$, EVPA=$77\pm5^\circ$\\
Total Counts: 124256
}\\

$\Delta\log L$ &
$8.7$ &
$10.2$ \\

PD variation significance &
$3.15~\sigma$ &
$3.53~\sigma$ \\

\hline

\textbf{Overlapping Model} & & \\

Head &
\makecell[c]{
PD=$45\pm15\%$, EVPA=$3.7\pm11^\circ$\\
Detection: $1.66~\sigma$\\
Counts: 710
}
&
\makecell[c]{
PD=$29\pm12\%$, EVPA=$-1.6\pm12^\circ$\\
Detection: $1.41~\sigma$\\
Counts: 706
}
\\

Spine &
\makecell[c]{
PD=$35\pm18\%$, EVPA=$-3.8\pm15^\circ$\\
Detection: $1.05~\sigma$\\
Counts: 1562
}
&
\makecell[c]{
PD=$27\pm14\%$, EVPA=$-4.8\pm15^\circ$\\
Detection: $1.11~\sigma$\\
Counts: 1554
}
\\

Sheath &
\makecell[c]{
PD=$10\pm20\%$, EVPA=$6.5\pm58^\circ$\\
Detection: $0.41~\sigma$\\
Counts: 2130
}
&
\makecell[c]{
PD=$10\pm16\%$, EVPA=$-2.4\pm47^\circ$\\
Detection: $~0.52\sigma$\\
Counts: 2119
}
\\

Photon Background &
\makecell[c]{
DU1: PD=$3\pm6\%$, EVPA=$-72\pm58^\circ$\\
DU2: PD=$3\pm6\%$, EVPA=$82\pm68^\circ$\\
DU3: PD=$5\pm6\%$, EVPA=$-9\pm39^\circ$\\
Total Counts: 32925
}&
\makecell[c]{
DU1: PD=$1\pm6\%$, EVPA=$44\pm177^\circ$\\
DU2: PD=$1\pm6\%$, EVPA=$-9\pm201^\circ$\\
DU3: PD=$5\pm6\%$, EVPA=$-9\pm40^\circ$\\
Total Counts: 31747
}
\\
Particle Background &
\makecell[c]{
DU1: PD=$2.3\pm0.7\%$, EVPA=$-86\pm9^\circ$\\
DU2: PD=$0.7\pm0.7\%$, EVPA=$54\pm29^\circ$\\
DU3: PD=$1.0\pm0.7\%$, EVPA=$76\pm20^\circ$\\
Total Counts: 130172
}& 
\makecell[c]{
DU1: PD=$2.2\pm0.7\%$, EVPA=$-79\pm9^\circ$\\
DU2: PD=$0.4\pm0.7\%$, EVPA=$36\pm48^\circ$\\
DU3: PD=$1.8\pm0.7\%$, EVPA=$74\pm12^\circ$\\
Total Counts: 124575
}
\\
$\Delta\log L$ &
$11.4$ &
$11.5$ \\

PD variation significance &
$3.81~\sigma$ &
$3.83~\sigma$ \\

\hline
\end{tabular}

\caption{Summary of the polarization fit results. For each component, the fitted PD, EVPA, polarization detection significance, and effective number of counts are listed. $\Delta\log L$ refers to the log-likelihood difference between the three-component model and the single-zone fit, corrected for the four additional free parameters (Stokes $q$ and $u$ for each extra region) of the latter. The PD variation significance is calculated following the procedure described in Section~\ref{data analysis}. Here we emphasize the substantial overall significance improvement of the three-component model over the single-zone model, rather than the lower significance for polarization detection or difference between individual regions.}

\label{result_table}
\end{table*}

\section{discussion}\label{discussion}
\subsection{Turbulence Modeling}\label{turbulence modeling}
Synchrotron theory implies a maximum polarization $\Pi_0=\Gamma / (\Gamma+2/3)$ for power-law electrons producing photon index $\Gamma$. With our $\Gamma=1.54$ we expect $PD<\Pi_0=69.8\%$. Turbulence or unmodeled macroscopic variations in the field will depolarize the source, and therefore the measured polarization constrains the ratio between ordered and turbulent fields.

\begin{figure*}
    \centering
    \includegraphics[width=0.35\linewidth]{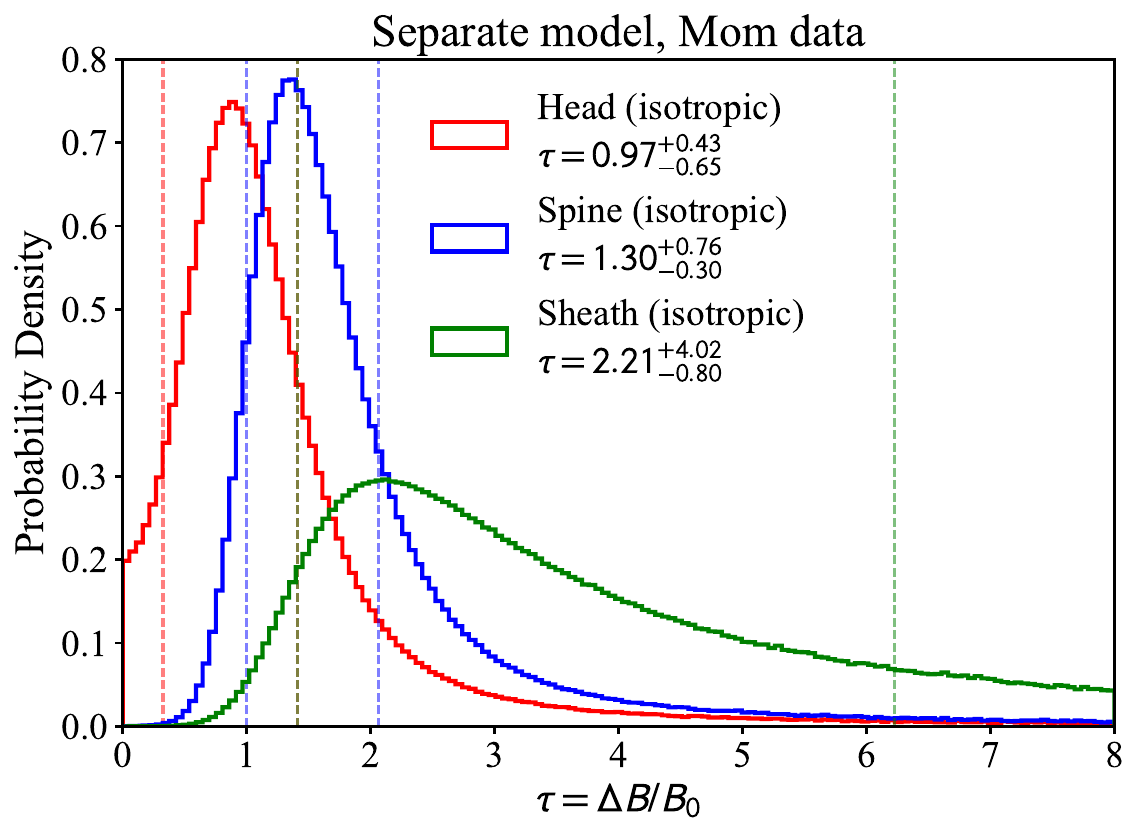}
    \includegraphics[width=0.35\linewidth]{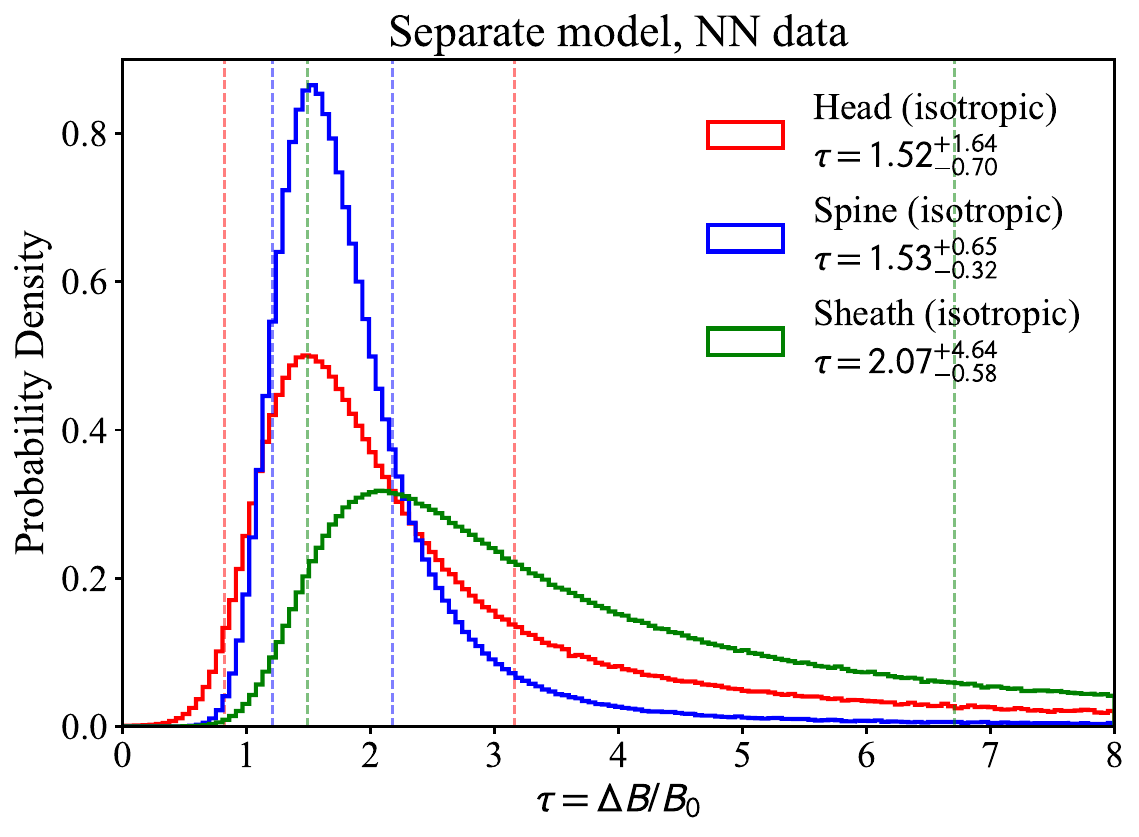}
    \includegraphics[width=0.35\linewidth]{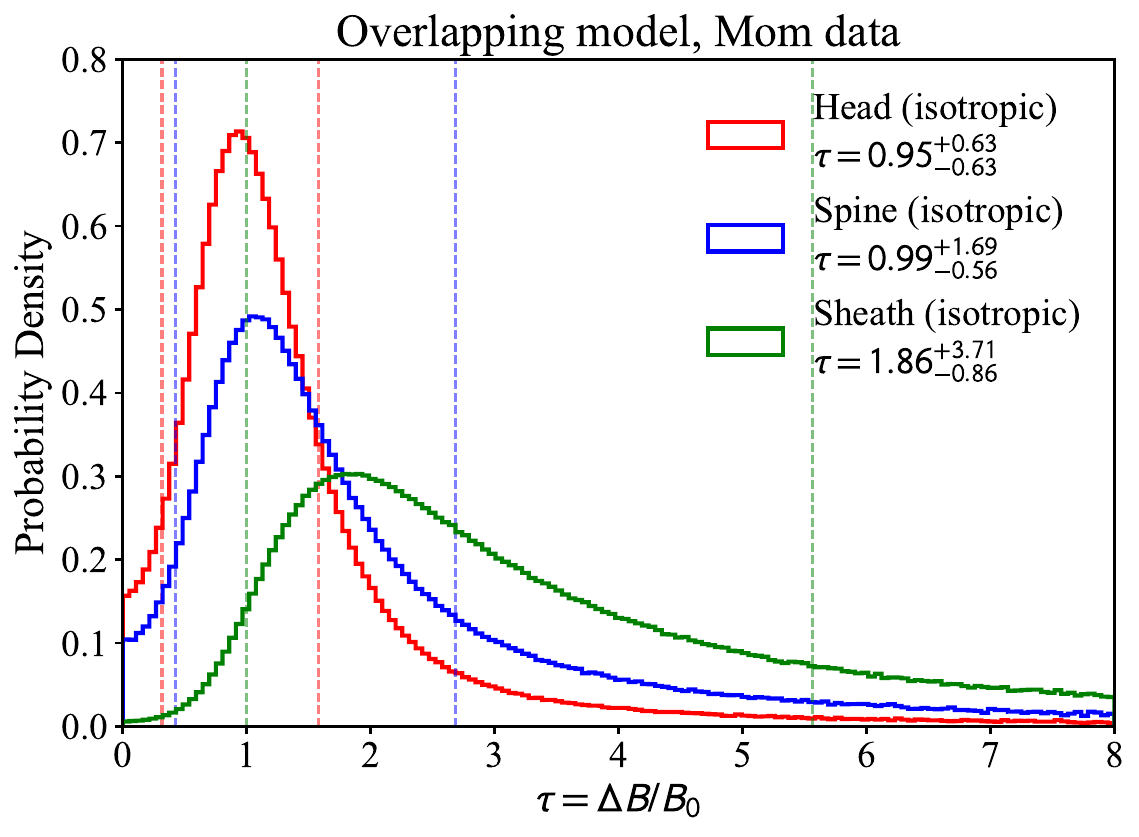}
    \includegraphics[width=0.35\linewidth]{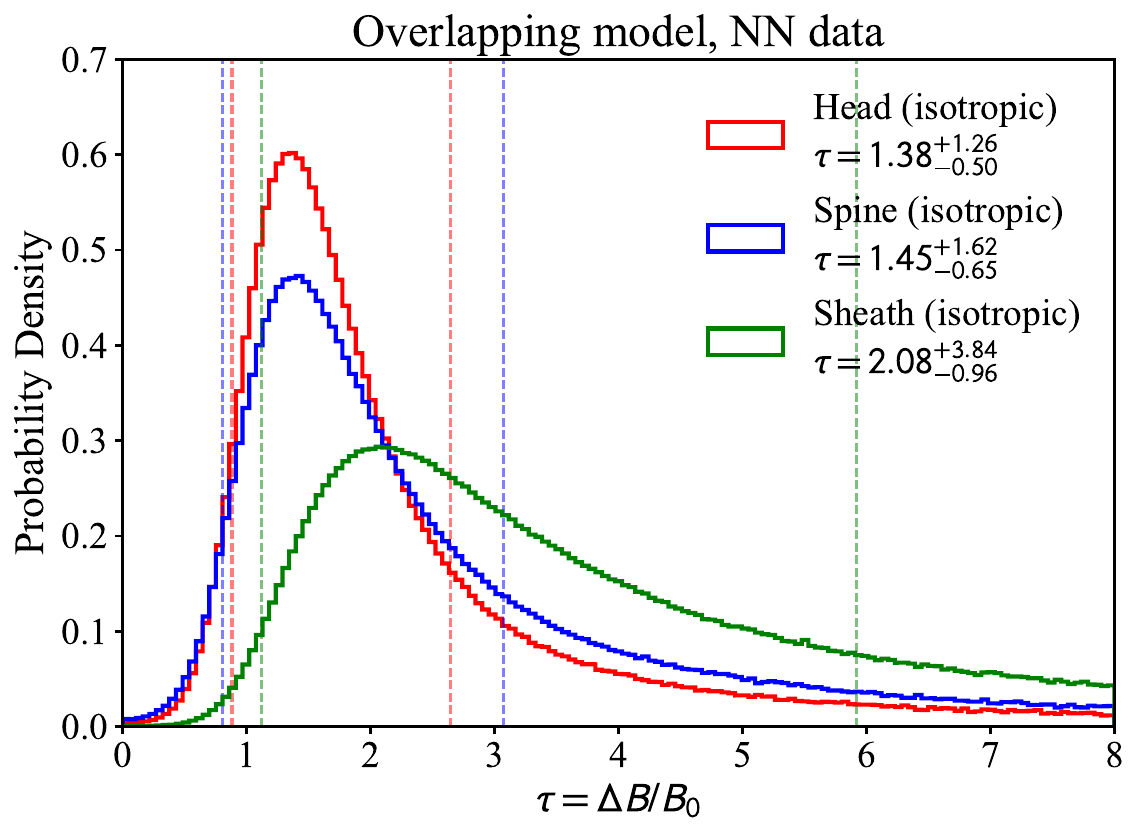}
\caption{The probability density distribution of the modeled turbulent ratio $\tau$ for different fits. Red, blue and green lines indicate the head, spine and sheath turbulence ratios, respectively. The vertical dashed lines indicate the $1~\sigma$ range.}
    \label{turbulence distribution}
\end{figure*}

\cite{Dinsmore2026_lighthouse} gives analytic expressions for the turbulent ratio $\tau=\delta
B/B_0$ , where $B_0$ is the unperturbed field strength. For Gaussian-distributed isotropic turbulent fluctuations $\delta B$, 
\begin{equation}
    \tau=\sqrt{\frac{3}{2}\frac{\Pi_0-PD}{PD}}\sin~i,
\end{equation} 
where $i$ is the angle between the magnetic field and line of sight (LoS). Since our observed EVPA follows the EXJ axis, we can assume the mean inclination of the jet precession axis $\sim79^{\circ}$ \citep{Margon1984} for the head. The spine and sheath however have a significant divergence of $\sim14^{\circ}$ from this axis as measured from the \textit{Chandra} image. If we assume a local magnetic field parallel to the region boundary, $i$ will vary through these zones as $i=\arccos[\sin7^\circ \cos11^\circ\cos\phi+\cos7^\circ\sin11^\circ]$, where $\phi$ is uniformly distributed between 0 and $2\pi$. For the predicted $\tau$ we integrate across the spine/sheath regions, using this varying $i$ (although the results are only slightly changed from those using the mean jet axis value). The $\tau$ posterior PDFs obtained from the various fits are shown in Fig.~\ref{turbulence distribution}, calculated in the same manner as \cite{Dinsmore2026_lighthouse}.

The typical values of the head and spine turbulence ratios are $\tau \lesssim1.5$, while for the sheath they are $\gtrsim1.8$, and the sheath $\tau$ has a more extended distribution than the other two. Drawing pairs of random samples from the posterior distributions, we find that the sheath $\tau$ is higher than that of the head or spine at the $1-1.8\sigma$ level in all the fits.

\subsection{Physical Interpretation}\label{turbulent sheath}

Our basic result is that the EXJ-aligned magnetic field inferred by \cite{Kaaret2024_ixpe} persists throughout the \textit{IXPE}-covered base of the EXJ, starting at the termination shock and persisting into the initial post-shock flow. The EVPA measurement is strongest for the spine, even though we have comparable photon counts from the surrounding sheath. This implies that the large scale field imposed at the termination shock persists downstream, but that the polarization is suppressed toward the EXJ boundary. This argues against another possible scenario in which unmodeled macroscopic changes in the ordered magnetic-field direction depolarize the boundary. One example of such a configuration is a hoop-stressed, confined jet spine with a transverse magnetic field embedded in a sheath with shear-induced field alignment parallel to the jet.

The decreased sheath PD is naturally interpreted as an increase in magnetic turbulence generated at the jet boundary, with $\tau$ roughly $2\times$ larger than in the head or spine. Astrophysical jets display several magnetohydrodynamic (MHD) instabilities \citep{schekochihin2022mhd_trubulence_review}. Of these, the Kelvin–Helmholtz (KH) instability driven by the velocity shear between the EXJ flow and the surrounding W50 medium provides a natural source of induced turbulence, and the associated mixing and entrainment between the two flows can further enhance the disorder at the jet boundary. Note that turbulence growth is expected for high Alfvén Mach number $M_{\rm A}=v/v_{\rm A}\gg1$, where $v_{\rm A}=B/\sqrt{4\pi\rho}$ is the Alfvén speed, because magnetic tension is insufficient to suppress the growth of shear instabilities \citep{frank1995_2D_KH,zhang2009_3D_KH}. We can adopt literature estimates for the EXJ post-shock flow: $n_\mathrm{H}\sim0.2~{\rm cm^{-3}}$ \citep{Sunyaev2025_erosita}, $v\sim0.08c$ \citep{HESS2024} and $B\sim12~\mu{\rm G}$ \citep{Safi-Harb_2022}. These give $M_\mathrm{A}=1300\times(\frac{v}{0.08~c})(\frac{B}{12~\mu G})^2(\frac{0.2~\mathrm{cm^{-3}}}{n_\mathrm{H}})^2$. Thus, even with the enhanced $\delta B/B_0\gtrsim2$ turbulent field of the sheath, we expect strong KH growth at the sheath/W50 interface.


\section{Conclusions}\label{conclusion}

We present a spatially resolved X-ray polarimetric study of the eastern EXJ in SS~433/W50 using \textit{IXPE}. Using the maximum-likelihood fitting framework of \texttt{LeakageLib} with the spatial and spectral information from \textit{Chandra} observations, we obtained sufficient sensitivity increase over the base-line \texttt{PCUBE} and spectropolarimetric analyses in \citep{Kaaret2024_ixpe}, to allow fitting to several EXJ components. 

From the \textit{Chandra} images, we defined three EXJ regions, the head, spine and sheath. These were modeled both as simple aperture sources and with an overlapping spatial model. The data were fit both with standard Moments-derived event parameters and with those inferred from Neural-Net processing. In all cases we find $>3~\sigma$ improvements to the global fit using the three-zone models. Separately, the head and spine PDs are higher than the sheath value, while the EVPAs of all three regions are consistent with $\sim0^{\circ}$, indicating that the average magnetic field is persistently aligned with the jet axis. We can connect the fitted PDs with the level of turbulence in the individual zones. For isotropic turbulence we infer a sheath turbulent ratio $\tau=\delta B / B$ approximately twice that of the head and spine. This can naturally be interpreted as shear-driven KH instability growth of turbulence in the sheath boundary layer, although the difference is presently at the $1-1.8\sigma$ level, depending on the model for the distribution of sheath flux.

Given the low surface brightness of the eastern EXJ and the substantial astrophysical and local particle/photon backgrounds, detailed event weights were crucial for this measurement. As \textit{IXPE} observes ever fainter sources, such techniques become increasingly important. Analysis of SS~433's western EXJ could benefit from similar event-weighted likelihood analysis. Still, higher sensitivity and spatial resolution will likely be needed to fully explore the polarization structure of SS~433's extended jets and relate the magnetic field and turbulence to the extreme particle acceleration evidently associated with these structures. 


\vspace{0.5cm}

{\bf Acknowledgments --} We thank Phil Kaaret and the anonymous referee for a careful reading of the manuscript and comments which improved the final version. The Imaging X-ray Polarimetry Explorer (IXPE) is a joint US and Italian mission. The US contribution is supported by the National Aeronautics and Space Administration (NASA) and led and managed by its Marshall Space Flight Center (MSFC), with industry partner Ball Aerospace (contract NNM15AA18C). The Italian contribution is supported by the Italian Space Agency (Agenzia Spaziale Italiana, ASI) through contract ASI-OHBI-2017-12-I.0, agreements ASI-INAF-201712-H0 and ASI-INFN-2017.13-H0, and its Space Science Data Center (SSDC), and by the Istituto Nazionale di Astrofisica (INAF) and the Istituto Nazionale di Fisica Nucleare (INFN) in Italy. 
This work was supported in part by NASA grants 80NSSC25K0277 and 80NSSC25K0278.

\smallskip
{\it Facilities}: IXPE, Chandra, XMM-Newton.
\smallskip

{\it Software}: LeakageLib (Dinsmore 2025), IXPEOBSSIM (Baldini et al. 2022), ASTROPY (Astropy Collaboration et al. 2013), XSPEC (Arnaud 1996) 

\bibliographystyle{aasjournal}
\bibliography{ref}

\end{document}